\documentclass[pre,10pt,onecolumn,preprintnumbers,superscriptaddress]{revtex4-2}

\usepackage{amsmath}
\usepackage{hyperref}
\usepackage{graphicx}
\usepackage{amsfonts}
\usepackage{amsthm}
\usepackage{cases}
\usepackage{bm}

\usepackage{color}
\definecolor{Blue}{rgb}{0.00, 0.00, 1.00}
\definecolor{Red}{rgb}{1.00, 0.00, 0.00}
\definecolor{Green}{rgb}{0.00, 0.70, 0.00}

\hypersetup{
    colorlinks=true,       
    linkcolor=red,          
    citecolor=blue,        
    filecolor=magenta,      
    urlcolor=cyan           
}

\newcommand{\beq}{\begin{equation}}
\newcommand{\eeq}{\end{equation}}
\newcommand{\beqn}{\begin{eqnarray}}
\newcommand{\eeqn}{\end{eqnarray}}

\newcommand{\dd}{\ensuremath{\mathrm d}}

\newcommand{\Deff}{D_\mathrm{eff}}

\newcommand{\eg}{{e.g., }}
\newcommand{\ie}{{i.e., }}
\newcommand{\brho}{{\overline{\rho}}}

\begin{document}

\title{Single-file diffusion in spatially inhomogeneous systems}
\author{Benjamin \surname{Sorkin}}
\affiliation{School of Chemistry and Center for Physics and Chemistry of Living Systems, Tel Aviv University, 69978 Tel Aviv, Israel}
\author{David S. \surname{Dean}}
\email{david.dean@u-bordeaux.fr}
\affiliation{Univ. Bordeaux, CNRS, LOMA, UMR 5798, F-33400, Talence, France.}
\affiliation{Team MONC, INRIA Bordeaux Sud Ouest, CNRS UMR 5251, Bordeaux INP, Univ. Bordeaux, F-33400, Talence, France.}
\date{\today}
\begin{abstract} 
We study the effect of spatially-varying potential and diffusivity on the dispersion of a tracer particle in single-file diffusion. Non-interacting particles in such a system exhibit normal diffusion at late times, which is characterised by an effective diffusion constant $D_\mathrm{eff}$. Here we demonstrate the physically appealing result that the dispersion of single-file tracers in this system has the same long-time behavior as that for Brownian particles in a spatially-homogeneous system with constant diffusivity $D_\mathrm{eff}$. Our results are based on a late-time analysis of the Fokker-Planck equation, motivated by the mathematical theory of homogenization. The findings are confirmed by numerical simulations for both annealed and quenched initial conditions.
\end{abstract}

\maketitle
\section{Introduction}\label{intro}

Transport properties in heterogeneous media can often be quantified on large length and timescales in terms of effective transport coefficients. For example, consider a Brownian particle diffusing in one dimension with a local diffusivity $D(x)$ subjected to an external potential $\phi(x)$, both of which vary periodically with the same period $\ell$. The Fokker-Planck equation (FPE) describing the evolution of the particle's probability density function $p(x,t)$ is  
\begin{equation}
\frac{\partial p(x,t)}{\partial t} =\hat{H}p(x,t),\qquad\hat{H}\equiv\frac{\partial}{\partial x}\left[D(x)\left(\beta \frac{d\phi(x)}{dx} + \frac{\partial }{\partial x}\right)\right],\label{de}
\end{equation}
where $\beta = (k_\mathrm{B}T)^{-1}$ is the inverse temperature. The mean-squared displacement (MSD) of the particle at late times behaves as
\begin{equation}
\langle [X(t)-X(0)]^2\rangle \simeq 2\Deff t,\label{defde}
\end{equation}
where $\Deff$ is called the effective (or, late-time) diffusion constant. 

The first derivation of $\Deff$ is attributed to Lifson and Jackson~\cite{LJ62}. Given that Eq.~\eqref{de} cannot be solved analytically in general, their argument was based on calculating the mean first-passage time to some large distance from the starting point. They argued that at large lengthscales and late times, the process is still diffusive, and can be described by a simple Brownian motion with an effective diffusion constant, $\Deff$. This means that the particle obeys the effective FPE
\begin{equation}
\frac{\partial p(x,t)}{\partial t} =\hat{H}_\mathrm{eff}p(x,t),\qquad\hat{H}_\mathrm{eff}\equiv\Deff\frac{\partial^2}{\partial x^2}\label{def}.
\end{equation}
Upon computing the mean first-passage time for Eq.~\eqref{de}, a comparison with that of Eq.~\eqref{def} yields the Lifson-Jackson formula,
\begin{equation}
\Deff = \frac{\ell^2}{\left[\int_0^\ell dx e^{\beta\phi(x)}/D(x)\right]\left[\int_0^\ell dxe^{-\beta\phi(x)}\right]}.\label{lj}
\end{equation}

Various direct computations of $\Deff$~\cite{DDH94,DDH07,GD15,GD215} from the definition of the MSD in Eq.~\eqref{defde} reassuringly agree with Eq.~\eqref{lj}, as one might expect from physical intuition. The Lifson-Jackson formula has been verified in a number of experimental settings, in particular, and of importance for our current study, the dependence on variations in the potential which can be modulated in the context of optical trapping \cite{HDS12,EZH13,ZBO22}.
It can further be shown that the late-time asymptotic correction to Eq.~\eqref{defde} is a constant~\cite{DO14}, and so the decay of the time-dependent diffusion coefficient is algebraic ($\sim1/t$) rather than exponential. One should bear in mind that the mean first-passage time argument occasionally breaks down as in, \eg the Sinai model~\cite{SI82} for diffusion in a Brownian random potential, where the tracer's motion is subdiffusive~\cite{CD98}. 

At the same time, the problem of single-file diffusion (SFD) has received much attention from both a theoretical~\cite{AP78,ADLV08,BD04,PM08,PS02,KM12,DG09,HA65,SP70,DGL85,RI77,AR83,LI83,LA08,KMS14,LB13,KOL03,RKH98} and, more recently, experimental point of view~\cite{KR92,WBL00,LKB04,LBE09,YYKJOCHK10,EHZCBDFJLHCPE13}. SFD occurs  in one-dimensional systems where particles cannot cross each other due to a hard-core repulsion term in their interactions. One way of treating this quantitatively is by ignoring the hard-core interaction, and relabelling the particles when they cross each other. The processes so generated are referred to as `elastically-colliding stochastic processes'~\cite{DGL85}. 
Within this picture, a tracer particle which was initially at the origin must have the same number of particles to its left at all times. If the density profile of the particles at time $t$ is given by $\rho(x,t)$, we get the constraint
\begin{equation}
\int_{-\infty}^{Y(t)} dx\rho(x,t) = \int_{-\infty}^{0} dx\rho(x,0),\label{link}
\end{equation}
where $Y(t)$ is the tracer's (stochastic) position at time $t$. Using this observation, the MSD of the tracer particle can be computed directly at late times by appealing to the central limit theorem (CLT); a method we will use in this paper. Additional methods, useful for the case of pure Brownian motion (\ie no interactions except for the hard-core repulsion), include the Bethe ansatz for the full joint probability density function~\cite{RKH98}, and the macroscopic fluctuation theory~\cite{KMS14,BDGJL01,BDGJL05,BDGJ06}.

There are two types of averages in play when treating SFD. The first average is over the thermal noise acting upon each of the individual particles, which we will denote by $\langle \cdots\rangle$. The second average is over the initial conditions, that is, the position at which each particle starts, which we will denote by $\overline{\cdots}$. Consequently, there are two possible definitions for the MSD~\cite{DG09}\,---\,the so-called \emph{annealed} variance
\begin{subequations}
\begin{equation}
\langle Y^2(t)\rangle_\mathrm{ac} = \overline{\langle Y^2(t)\rangle} - \overline{\langle Y(t)\rangle}\; \overline{\langle Y(t)\rangle},
\end{equation}
and the \emph{quenched} variance 
\begin{equation}
\langle Y^2(t)\rangle_\mathrm{qc} = \overline{\langle Y^2(t)\rangle} - \overline{\langle Y(t)\rangle\; \langle Y(t)\rangle}.
\end{equation}
\end{subequations}
The initial conditions are assumed to be periodic with $\ell_0$ and have an average particle density denoted by $\brho$. 

For free Brownian motion, numerous authors (\eg Refs.~\cite{HA65,AP78,KMS14}) have shown that
\begin{subequations}
\begin{equation}
\langle Y^2(t)\rangle_\mathrm{ac}= \frac{2}{\brho}\sqrt{\frac{Dt}{\pi}}\label{sfanon}
\end{equation}
and
\begin{equation}
\langle Y^2(t)\rangle_\mathrm{qc}=\frac{\sqrt2}{\brho}\sqrt{\frac{Dt}{\pi}},\label{sfqnon}
\end{equation}
\end{subequations}
where $D$ is the {\em bare} diffusion constant of the particles. The subdiffusive behavior is a consequence of the hard-core interaction. We also see that the annealed MSD is larger than the quenched one as the fluctuations in the initial conditions of the former lead to additional fluctuations in the displacement of the tracer.

We now come to the main question of this paper. When considering systems with a periodically varying potential and diffusivity (instead of pure Brownian diffusion), can one simply replace the bare diffusion coefficient $D$ in Eqs.~\eqref{sfanon} and~\eqref{sfqnon} with $\Deff$ at late times? That is, write
\begin{subequations}
\begin{equation}
\langle Y^2(t)\rangle_\mathrm{ac}= \frac{2}{\brho}\sqrt{\frac{\Deff t}{\pi}}\label{sfa}
\end{equation}
and
\begin{equation}
\langle Y^2(t)\rangle_\mathrm{qc}=\frac{\sqrt2}{\brho}\sqrt{\frac{\Deff t}{\pi}}.\label{sfq}
\end{equation}
\end{subequations}
Simulations of SFD under a periodic potential~\cite{TM06,LRM19} support Eq.~\eqref{sfa} (and furthermore this idea also works when a constant applied force is added to that produced by the periodic potential). Within liquid theory, a very similar expression to Eq.~\eqref{sfa} for a tracer's MSD has been derived for ensembles of SFD particles with pairwise interactions~\cite{KOL03}. A general inequality relation with the entropy has been derived recently, becoming particularly tight for SFD~\cite{SDA22}. As far as we know there have been no numerical studies of the effect of locally varying diffusivity in SFD. 

Here we will show analytically that Eqs.~\eqref{sfa} and~\eqref{sfq} do indeed hold at late times. To do so, we will exploit Eq.~\eqref{link}\,---\,the link between SFD and the effusion problem for non-interacting Brownian particles described by the FPE, Eq.~\eqref{de}. 
The effusion problem, which counts the number of crossings from left (right) to right (left) at the origin, can be solved by a late-time asymptotic analysis of the FPE for a single particle. The method we use is closely related to the mathematical theory of homogenization~\cite{PS08,NG89,AL92}. 
The fact that the transport coefficient we compute is that of a tracer makes our prediction particularly relevant for optical trapping experiments~\cite{EHZCBDFJLHCPE13} where a spatially varying potential can be generated. For example, theoretical results such as Eq.~\eqref{lj} have been confirmed in experiments on single colloids~\cite{MLT13,SMLT17}. 

The paper is organised as follows: In Sec.~\ref{slink}, we recall how the MSD of a tracer particle can be described in terms of an effusion problem. There we give an explicit expression for the MSD in terms of the solution to the single-particle FPE. In Sec.~\ref{latetime}, we carry out an asymptotic analysis of these results to prove Eqs.~\eqref{sfa} and~\eqref{sfq}. In Sec.~\ref{num}, we confirm our analytical predictions using the results of a numerical simulation. Finally, we conclude and discuss possible applications and extensions of our results in Sec.~\ref{conc}.

\section{Link between single-file diffusion and effusion}\label{slink}

Here we recall the basic arguments given in Refs.~\cite{DGL85,DMS23}, which link the displacement of a single-file tracer with the effusion of non-interacting Brownian processes to the left and right of the origin. The key starting point for our analysis is Eq.~\eqref{link}, which can be rewritten as 
\begin{equation}
\int_{0}^{Y(t)} dx\rho(x,t) = \int_{-\infty}^{0} dx\rho(x,0)-\int_{-\infty}^{0} dx \rho(x,t). \label{link2}
\end{equation}

We first consider the left-hand side. If we assume that the typical value of $Y(t)$ grows as a function of $t$, then at sufficiently late times, we will have $|Y(t)| \gg\ell$. Thus, the integral on the left-hand side can be approximated, using the CLT, by
\begin{equation}
\int_{0}^{Y(t)} dx\rho(x,t) \simeq \brho Y(t),\label{asym}
\end{equation}
with corrections ${\cal O}(\sqrt{\brho|Y(t)|})$. 

The term on the right-hand side is linked to an effusion problem as follows. Suppose the particles initially to the right and left of the origin are two different particle types (say, `left type' and `right type'). The two particle types are non-interacting, and both obey the same FPE. Denote their densities as $\rho_\mathrm{L}(x,t)$ and $\rho_\mathrm{R}(x,t)$. 
This implies that, in the beginning, $\rho_\mathrm{L}(x>0,0)=0$ and $\rho_\mathrm{R}(x<0,0)=0$, while the total density is simply $\rho(x,t)=\rho_\mathrm{L}(x,t)+\rho_\mathrm{R}(x,t)$.
We define $N_+(t)$ ($N_-(t)$) to be the number of particles that were initially to the left (right) at time $0$ and are on the right (left) at time $t$. Clearly,
\begin{subequations}
\label{cross}
\begin{eqnarray}
N_+(t) &=& \int_0^\infty dx\rho_\mathrm{L}(x,t),\\
N_-(t) &=& \int_{-\infty}^0 dx\rho_\mathrm{R}(x,t).
\end{eqnarray}
\end{subequations}
We impose conservation of the left-type particle number through
\begin{equation}
 \int_0^\infty  dx\rho_\mathrm{L}(x,t)+ \int_{-\infty}^0  dx\rho_\mathrm{L}(x,t)= \int_{-\infty}^0  dx\rho_\mathrm{L}(x,0). \label{lcons}
\end{equation}
Combining Eqs.~\eqref{cross} and~\eqref{lcons}, we relate the change in density to the number of crossings,
\begin{eqnarray}
N_+(t) - N_-(t) &=&\int_{-\infty}^0  dx \rho_\mathrm{L}(x,0)-\int_{-\infty}^0  dx\rho_\mathrm{L}(x,t)-\int_{-\infty}^0 dx\rho_\mathrm{R}(x,t)\nonumber\\
&=&\int_{-\infty}^0  dx\rho(x,0)-\int_{-\infty}^0  dx \rho(x,t).\label{effs}
\end{eqnarray}

Using the asymptotic Eq.~\eqref{asym} and the effusion-related Eq.~\eqref{effs}, Eq.~\eqref{link2} may be rewritten as
\begin{equation}
Y(t) =\frac{1}{\brho}\left[N_+(t) - N_-(t)\right].
\end{equation}
Thus we have established the link between SFD and two independent effusion proccesses. The random variables $N_+(t)$ and $N_-(t)$ are independent and identically distributed. Both the annealed and quenched variances (which are identical for the first moment) are given by
$\langle Y(t) \rangle_{\mathrm{a}}= \langle Y(t) \rangle_{\mathrm{q}}=0$ in the absence of bias.
Expanding and using the independence and identical distribution of $N_+(t)$ and $N_-(t)$, we can write the annealed and quenched variances solely in terms of moments of, say, $N_+(t)$,
\begin{subequations}
\begin{eqnarray}
\langle Y^2(t)\rangle_\mathrm{ac} &=&\frac{2}{\brho^2} \left(\overline{\langle N_+^2(t)\rangle} 
- \overline{\langle N_+(t)\rangle}\;\overline{\langle N_+(t)\rangle}\right)\equiv\frac{2}{\brho^2}\langle N_+^2(t)\rangle_\mathrm{ac},\label{y2ac}\\
\langle Y^2(t)\rangle_\mathrm{qc} &=&\frac{2}{\brho^2}\left(\overline{\langle N_+^2(t)\rangle}- \overline{\langle N_+(t)\rangle\;\langle N_+(t)\rangle}\right)\equiv
\frac{2}{\brho^2}\langle N_+^2(t)\rangle_\mathrm{qc}. \label{y2qc}
\end{eqnarray}
\end{subequations}
Therefore, the MSD is given in terms of a single-sided effusion problem as studied in Ref.~\cite{DMS23}.

Now we examine the effusion problem more closely. The value of $N_+(t)$ can be written as
\begin{equation}
N_+(t)= \sum_{m=1}^M\Theta(X_m(t)),
\end{equation}
where $\Theta(x)$ is the Heaviside theta function, $X_m(t)$ is the diffusion process described by the FPE (Eq.~\eqref{de}), and $M$ is the number of particles that started to the left of the origin. 
We proceed by taking the averages over noise and initial conditions explicitly.
If $p(x,t|x_0,0)$ is the solution of Eq.~\eqref{de} (for the initial condition $p(x,0|x_0,0)=\delta(x-x_0)$), the average value of $N_+(t)$ over the thermal noise is
\begin{equation}
\langle N_+(t)\rangle=\sum_{m=1}^M\int_0^\infty dxp(x,t|x_{0m};0),\label{Npbdel}
\end{equation}
where $\{x_{0m}\}$ is a list of the particles' initial positions that are to the left of the origin. 
Suppose the particles' positions are drawn from a periodic initial distribution $P_0(x)$, with a wavelength $\ell_0$ which does not have to be the same as that of the potential or diffusivity (being $\ell$).
We define it to be normalized within one cell, $\int_0^{\ell_0} dxP_0(x)=1$, so
the initial density of particles to the left of the origin is then $\overline{\rho_\mathrm{L}(x)}=(\ell_0 M/L)P_0(x)$, where $L$ is the (large) size of the left half of the box.
Thus, the average value of $\langle N_+(t)\rangle$ over the initial positions is
\begin{eqnarray}
\overline{\langle N_+(t)\rangle}&=&\int_{-L}^0  dx_0 \overline{\rho_\mathrm{L}(x_0)}\int_0^\infty dx  p(x,t|x_{0m};0)=M\int_{-L}^0dx_{0} \frac{\ell_0}{L}P_0(x_{0})\int_0^\infty dx  p(x,t|x_{0m};0)\nonumber\\
&=&\brho\ell_0\int_{-\infty}^0dx_0 P_0(x_0)\int_0^\infty dx p(x,t|x_0,0),\label{navg}
\end{eqnarray}
where we have taken the limits $M\to\infty$ and $L\to\infty$, while keeping $\brho= M/L$ fixed.

Finally, we compute the quenched and annealed variances. Since the particles' Brownian motions (\ie the thermal noise values) are independent, we have
\begin{subequations}
\begin{equation}
    \langle N^2_+(t)\rangle= \sum_{m=1}^M   \langle \Theta(X_m(t))\rangle + \sum_{m=1}^M\sum_{n\ne m}^M   \langle \Theta(X_m(t))\rangle \langle \Theta(X_n(t))\rangle,\label{npmmnt}
\end{equation}
and thus
\begin{equation}
    \langle N^2_+(t)\rangle-\langle N_+(t)\rangle^2=\sum_{m=1}^M \langle\Theta(X_m(t))\rangle - \sum_{m=1}^M\langle \Theta(X_m(t))\rangle^2,\label{npvar}
\end{equation}
\end{subequations}
where we have used $\Theta^2=\Theta$. Since the particle positions are drawn from the initial distribution independently (unlike the case we consider in Appendix~\ref{apa}; see below), averaging Eq.~\eqref{npmmnt} over the initial positions yields
\begin{equation}
\overline{\langle N^2_+(t)\rangle}= \int_{-\infty}^0 dx_0 \overline{\rho_\mathrm{L}(x)}\int_0^\infty dx  p(x,t|x_0,0)+ \overline{\langle N_+(t)\rangle}\;\overline{\langle N_+(t)\rangle}.
\end{equation}
This leads to the annealed variance
\begin{subequations}
\begin{equation}
\langle N^2_+(t)\rangle_\mathrm{ac} = \int_{-\infty}^0 dx_0  \overline{\rho_\mathrm{L}(x)} \int_0^\infty dx  p(x,t|x_0,0).\label{n2ac}
\end{equation}
On the other hand, we obtain the quenched variance by averaging Eq.~\eqref{npvar} over the initial conditions  
\begin{equation}
\langle N^2_+(t)\rangle_\mathrm{qc}=  \int_{-\infty}^0 dx_0 \overline{\rho_\mathrm{L}(x)}  \int_0^\infty dx p(x,t|x_0,0)-\int_{-\infty}^0 dx_0 \overline{\rho_\mathrm{L}(x)}\int_0^\infty dx p(x,t|x_0,0)\int_0^\infty dx'  p(x',t|x_0,0).\label{n2qc}
\end{equation}
\end{subequations}
The two averaging procedures above apply to the same physical system where all initial positions are independent, yielding two different statistics. However, the quenched average has the same mathematical expression as that for the annealed average, carried out on a system where the initial density fluctuations are suppressed~\cite{BJC22,DMS23}, in particular if the initial density is periodic or hyperuniform~\cite{TOR18}. We discuss this point Appendix~\ref{apa} and use it to simplify the simulation procedure in Sec.~\ref{num}.

\section{Late time analysis of the Fokker-Planck equation}
\label{latetime}

Looking at the expressions in Eqs.~\eqref{y2ac}, \eqref{n2ac}, \eqref{y2qc}, and~\eqref{n2qc}, we see that they depend on the solution $p(x,t|x_0,0)$ of the FPE, Eq.~\eqref{de}, at late times. However, they only depend on terms of the form (which we define as for later convenience)
\begin{equation}
\frac{f(x_0,t)}{P_\mathrm{B}(x_0)} =  \int_{-\infty}^\infty dx\ p(x,t|x_0,0)\Theta(x),\label{fkesol}
\end{equation}
where $P_\mathrm{B}(x)$ is the equilibrium Boltzmann factor,
\begin{subequations}
\begin{equation}
P_\mathrm{B}(x) =Z^{-1}e^{-\beta\phi(x)},\label{eqpbol}
\end{equation}
with the partition function 
\begin{equation}
Z = \int_0^\ell dx e^{-\beta \phi(x)}.
\end{equation}
\end{subequations}
It is easy to show that  $f(x,t)/P_\mathrm{B}(x)$ solves the adjoint FPE (or, the so-called Kolmogorov backward equation), $\partial (f/P_\mathrm{B})/\partial t=\hat{H}^\dagger (f/P_\mathrm{B})$, where $\hat{H}^\dagger$ is the adjoint operator of $\hat{H}$, and $\hat{H}$ as  written in Eq.~\eqref{de}~\cite{OKS03}. We then see that $f(x,t)$ obeys the same FPE (the Kolmogorov forward equation) as Eq.~\eqref{de},
\begin{subequations}
\begin{equation}
\frac{\partial f(x,t)}{\partial t} = \hat{H} f(x,t),\label{f1}
\end{equation}
with the initial condition
\begin{equation}
f(x,0) = \Theta(x)P_\mathrm{B}(x).\label{f1ic}
\end{equation}
\end{subequations}

We now take the Laplace transform of Eq.~\eqref{f1}, which yields the equation for $\tilde f(x,s) = \int_0^\infty dt\exp(-st) f(x,t)$,
\begin{equation}
\hat{H}\tilde f(x,s) = s \tilde f(x,s) - f(x,0).\label{fpels}
\end{equation}
The late time behavior of $f(x,t)$ can be extracted from the behavior of $\tilde f(x,s)$ at small $s$. To proceed, we follow the ideas of the mathematical theory of homogenization~\cite{PS08}. The diffusive motion is governed by some `macro-variable', $y=s^{1/2}x$, while the precise location within a cell is described by a `micro-variable', still denoted by $x$, and restricted to a single cell, $0<x<\ell$. Since we take the late-time (small-$s$) limit, we expect there to be a scale separation between $y$ and $x$. Note that the potential $\phi(x)$ and diffusivity $D(x)$ still depend only on the micro-variable $x$, and both are periodic in it with period $\ell$.

To formulate the above mathematically, we write the following expansion
\begin{equation}
\tilde f(x,s) = \frac{1}{s}\sum_{n=0}^\infty s^{\frac{n}{2}}F_n(x,\sqrt{s}x),\label{expans}
\end{equation}
where $\{F_n(x,y)\}$ are periodic in $x$ with period $\ell$.
Upon the  introduction of these  two spatial variables which are treated independently, the spatial derivatives in Eq.~\eqref{fpels} are modified to $(\partial/\partial x) \to (\partial/\partial x) + s^{1/2}\cdot(\partial/\partial y)$. We now insert the expansion of Eq.~\eqref{expans} in Eq.~\eqref{fpels}, and equate terms of the same order. To obtain an effective late-time diffusion equation of the form of Eq.~\eqref{def}, it turns out that we need the three lowest  orders.

Upon equating the leading order $\mathcal{O}(s^{-1})$ terms, we find
\begin{equation}
\frac{\partial}{\partial x}\left[D(x)\left(
\beta \frac{d\phi(x)}{dx}F_0(x,y) + \frac{\partial F_0(x,y)}{\partial x}\right)\right]=0.
\end{equation}
We see that we may assume separation of variables, and get a solution of the form
\begin{equation}
F_0(x,y) = P_\mathrm{B}(x)K_0(y),\label{F0}
\end{equation}
where $K_0(y)$ is a function of $y$ which we will be able to compute later, and $P_\mathrm{B}(x)$ is then the Boltzmann factor.
Clearly  $F_0(x,y)$ is periodic in $x$ with period length $\ell$. (Note that the initial distribution in the SFD process, $P_0(x)$, need not be the equilibrium Boltzmann factor, $P_\mathrm{B}$, or even have the same periodicity $\ell_0\ne\ell$. In fact, for the quenched case, we shall use a Dirac comb for $P_0(x)$ in Sec.~\ref{num}, as explained in Appendix~\ref{apa}.)

At the next order, $\mathcal{O}(s^{-1/2})$, we find
\begin{multline}
\frac{\partial}{\partial x}\left[D(x)\left( \beta \frac{d\phi(x)}{dx}F_1(x,y)+\frac{\partial F_1(x,y)}{\partial x} \right)\right] 
\\+\frac{\partial}{\partial y}\left[D(x)\left( \beta \frac{d\phi(x)}{dx}F_0(x,y)+ \frac{\partial F_0(x,y)}{\partial x} \right)\right]+\frac{\partial}{\partial x}\left[D(x)\frac{\partial F_0(x,y)}{\partial y}\right]=0.
\end{multline}
Now, using Eq.~\eqref{F0}, and again making the separation of variables ansatz,
\begin{equation}
F_1(x,y) = P_1(x)\frac{dK_0(y)}{dy},\label{F1ans}
\end{equation}
we find the equation for $P_1(x)$,
\begin{equation}
\frac{d}{d x}\left[D(x)\left( \beta \frac{d\phi(x)}{dx}P_1(x)+\frac{dP_1(x)}{dx} +P_\mathrm{B}(x)\right)\right]=0.\label{F1}
\end{equation}

The last $\mathcal{O}(1)$ terms give
\begin{multline}
\frac{\partial}{\partial x}\left[D(x)\left( \beta \frac{d\phi(x)}{dx}F_2(x,y)+\frac{\partial F_2(x,y)}{\partial x}\right)\right]
\\+\frac{\partial}{\partial y}\left[D(x)\left( \beta \frac{d\phi(x)}{dx}F_1(x,y)+\frac{\partial F_1(x,y)}{\partial x}\right)\right]
+\frac{\partial}{\partial x}\left[D(x)\frac{\partial F_1(x,y)}{\partial y}\right] \\+D(x)\frac{\partial^2 F_0(x,y)}{\partial y^2} = F_0(x,y) - \Theta(y)P_0(x). \label{thrdord}
\end{multline}
Here, we replace $\Theta(x)\to\Theta(y)$ since $y$ is the {\em macro-variable}. Inserting the ansatze of Eqs.~\eqref{F0} and~\eqref{F1ans}, and utilizing the periodicity of the solutions in $x$ to integrate Eq.~\eqref{thrdord} between $0$ and $\ell$ (thus canceling complete differentials), we find the equation for $K_0(y)$, 
\begin{equation}
\Deff\frac{\partial^2 K_0(y)}{\partial y^2}
= K_0(y) - \Theta(y).\label{K0}
\end{equation}
Equation~\eqref{K0} is of the same form as Eq.~\eqref{def} (written in Laplace space), where we identified the effective diffusion constant
\begin{equation}
\Deff=\int_0^\ell dx D(x)\left[\beta \frac{d\phi(x)}{dx}P_1(x) +\frac{dP_1(x)}{\partial x} + P_\mathrm{B}(x)\right].\label{defa}
\end{equation}
The solution to Eq.~\eqref{K0} is, using the required smoothness conditions, 
\begin{equation}
K_0(y) = \begin{cases}
1-\frac{1}{2}\exp\left(-y/\sqrt{\Deff}\right), & y>0,\\
\frac{1}{2}\exp\left(y/\sqrt{\Deff}\right). & y<0.
\end{cases}\label{caseK0}
\end{equation}

Prior to proceeding, we first find the explicit formula of $\Deff$ (which can only be  explicitly computed in one dimension~\cite{PS08}.) Integrating Eq.~\eqref{F1}, we find that 
\begin{equation}
 D(x)\left( \beta \frac{d\phi(x)}{dx}P_1(x)+\frac{dP_1(x)}{dx} +P_\mathrm{B}(x)\right)=C,
\end{equation}
where $C$ is a constant. Comparing it with Eq.~\eqref{defa}, we identify $\Deff = \ell C$. Solving for $P_1(x)$ and using its periodicity gives
\begin{equation}
C = \frac{\ell}{Z \left[\int_0^\ell dx e^{\beta\phi(x)}/D(x)\right]}.
\end{equation}
This then yields the Lifson-Jackson formula for the effective diffusion constant $\Deff$ of a Brownian particle, Eq.~\eqref{lj}.

Using these late-time solutions, we first compute the annealed variance. Note that we may rewrite Eq.~\eqref{n2ac} more compactly in terms of $f$ as
\begin{equation}
\langle N_+^2(t) \rangle_\mathrm{ac}= \brho \ell_0\int_{-\infty}^\infty dx_0\Theta(-x_0)P_0(x_0)\frac{f(x_0,t)}{P_\mathrm{B}(x_0)} .\label{anp}
\end{equation}
Upon combining the $y<0$ case of Eq.~\eqref{caseK0}, Eq.~\eqref{F0}, and Eq.~\eqref{expans}, we find, to leading order in small $s$ (corresponding to late times), 
\begin{equation}
\tilde{f}(x,s) \simeq \frac{1}{2s}P_\mathrm{B}(x)\exp\left(\sqrt{\frac{s}{\Deff}}x\right),
\end{equation}
for $y<0$, and see that it satisfies the initial conditions Eq.~\eqref{f1ic}.  Thus, Eq.~\eqref{anp} becomes, 
\begin{equation}
\langle \tilde{N}^2_+(s)\rangle_\mathrm{ac} = \frac{\brho\ell_0}{2s}\int_{-\infty}^0 dx_0 P_0(x_0)\exp\left(\sqrt{\frac{s}{\Deff}}x_0\right).\label{n2acsimp}
\end{equation}
Here we need to use the fact that $P_0(x)$ is periodic, so we can express it as a Fourier series,
\begin{equation}
P_0(x) = \frac{1}{\ell_0}\left[1+ \sum_{n\neq 0}\exp\left(\frac{2\pi  i n x}{\ell_0}\right) \int_0^\ell dx' P_0(x') \exp\left(-\frac{2\pi  i n x'}{\ell_0}\right) \right].\label{p0arg}
\end{equation}
Note that the terms coming from the non-zero Fourier modes $k_n = 2\pi n/\ell_0$ integrate as
\begin{subequations}
\label{fsbigs}
\begin{equation}
\int_{-\infty}^0 dx_0 \exp\left(\sqrt{\frac{s}{\Deff}}x_0+ik_nx_0\right)= \frac{1}{i k_n+\sqrt{s/\Deff}},
\end{equation}
while for $n=0$ we find
\begin{equation}
\int_{-\infty}^0 dx_0 \exp\left(\sqrt{\frac{s}{\Deff}}x_0\right)= \sqrt{\frac{\Deff}{s}}.\label{page}
\end{equation}
\end{subequations}
Therefore, at late times, the dominant term in Eq.~\eqref{n2acsimp} is  the one corresponding to $n=0$. Inverting the Laplace transform, we find
\begin{equation}
\langle N^2_+(t)\rangle_\mathrm{ac} = \brho\sqrt{\frac{\Deff t}{\pi}}.\label{nac}
\end{equation}
Using Eq.~\eqref{y2ac}, we obtain Eq.~\eqref{sfa} for the annealed version of SFD. 

To compute the quenched variance, we write Eq. (\ref{n2qc}) as
\begin{subequations}
\begin{equation}
\langle N_+^2(t)\rangle_\mathrm{qc}= \langle N_+^2(t)\rangle_\mathrm{ac}- \lambda(t),
\end{equation}
where, in terms of $f$,
\begin{equation}
\lambda(t)=\brho\ell_0\int_{-\infty}^\infty dx_0 \Theta(-x_0)P_0(x_0)\frac{f^2(x_0,t)}{P_\mathrm{B}^2(x_0)}.
\end{equation}
\end{subequations}
Since $\lambda(t)$ is not linear in $f$, it is not convenient to work in terms of its Laplace transform. Upon inserting Eq.~\eqref{caseK0} in Eq.~\eqref{F0}, an inverse Laplace transform of Eq.~\eqref{expans} gives for $x<0$, at late times,
\begin{equation}
f(x,t) \simeq \frac{1}{2}P_\mathrm{B}(x){\rm erfc}\left(-\frac{x}{2\sqrt{\Deff t}}\right).
\end{equation}
We then obtain
\begin{equation}
\lambda(t)=\frac{\brho\ell_0}{4}\int_{-\infty}^0 dx_0 P_0(x_0){\rm erfc}^2\left(-\frac{x_0}{2\sqrt{\Deff t}}\right),
\end{equation}
For large $t$, following the arguments of Eqs.~\eqref{p0arg} and~\eqref{fsbigs}, we may only keep the non-oscillatory component of $P_0(x_0)$, which is $1/\ell_0$. Therefore, we obtain
\begin{equation}
\lambda(t) = \brho\left(1-\frac{1}{\sqrt2}\right)\sqrt{\frac{\Deff t}{\pi}}.
\end{equation}
We thus find that
\begin{equation}
\langle N_+^2(t)\rangle_\mathrm{qc}=\brho\sqrt{\frac{\Deff t}{2\pi}}.\label{beldisc}
\end{equation}
Using Eq.~\eqref{y2qc}, we obtain Eq.~\eqref{sfq} for the quenched version of SFD, as well.

The above constitutes a proof that both the annealed and quenched MSDs of a tracer particle in SFD can be derived by considering purely Brownian particles but with an effective diffusion constant $\Deff$, given by the Lifson-Jackson formula. It takes fully into account the effects of periodically-varying potential and diffusivity at the level of large-time asymptotics. We find that both MSDs grow asymptotically as $\mathcal{O}[(\Deff t)^{1/2}/\brho]$. By keeping the next-order terms in the present Sec.~\ref{latetime}, one can show that the corrections to the MSD, associated with the late-time approximation, are of order $\mathcal{O}[\ell/\brho]$. A so-called `weak convergence theorem' was proven mathematically for general two-scale homogenization schemes~\cite{PS08,NG89,AL92}, rigorously justifying the replacement of the true $f(x,t)$ with leading order term in its asymptotic expansion. Note, however, that the homogenization-related  late-time corrections are smaller than the corrections associated with the application of the CLT (Eq.~\eqref{asym}), which are of order $\mathcal{O}[(\Deff t)^{1/4}/\brho^{3/2}]$.

\section{Numerical simulation}
\label{num}

In order to verify our analytical findings, we perform Langevin dynamics simulations in the overdamped limit. The It\^o stochastic differential equation corresponding to Eq.~\eqref{de} is~\cite{OKS03,LL07}
\begin{equation}
    \dd X(t)=\left(\left.\frac{dD(x)}{dx}\right|_{X(t)}-\beta D(X(t))\left.\frac{d\phi}{dx}\right|_{X(t)}\right)\dd t+\sqrt{2D(X(t))}\dd B(t).\label{langevin}
\end{equation}
When this is simulated we take $B(t)$ to be a discretized Brownian motion, with the discrete form of It\^o convention where its increments are Gaussian with $\langle\dd B(k\dd t)\dd B(l\dd t)\rangle=\delta_{kl}\dd t$ and zero mean. 

Below, we will consider the case of periodic potential, $\beta\phi(x)=E[1-\cos(2\pi x/\ell)]$, with constant diffusivity $D=D_0$. Equation~\eqref{lj} yields $\Deff(E)=D_0/I_0^2(E)$, where $I_0(z)$ is the zeroth modified Bessel function of the first kind. In addition to the parameters $E$, recall that the particle density $\brho$ (number of particles per period length $\ell$) is another control parameter in SFD problems.Thus, using Eqs.~\eqref{sfa} and~\eqref{sfq}, we predict
\begin{subequations}
\label{predPOT}
\begin{eqnarray}
    \langle Y^2(t;E,\rho)\rangle_\mathrm{ac}&=& \frac{2}{\brho I_0(E)}\sqrt{\frac{D_0 t}{\pi}},\\
    \langle Y^2(t;E,\rho)\rangle_\mathrm{qc}&=& \frac{\sqrt2}{\brho I_0(E)}\sqrt{\frac{D_0 t}{\pi}}.
\end{eqnarray}
\end{subequations}
This case was studied numerically also in Refs.~\cite{TM06,LRM19,SDA22}. 

The simulations were carried out as follows. We choose the unitless quantities $\ell=1$, $D_0=1/2$, and $\beta=1$ and vary the parameters $\brho$ and $E$. In all simulations, we spread $M=10^4\ell\brho$ particles within the box according to an initial distribution we explain below.
In each timestep, we ``sweep'' the system $M$ times. In each sweep, we pick a random particle, and move it according to Eq.~\eqref{langevin} with periodic boundary conditions in a box of size $10^4\ell$. We perform a total of $10^6$ timesteps of duration $\dd t=10^{-3}\ell^2/(2D_0)$. Each time a particle passes one of its neighbors, we relabel the relevant particles such that the `leftmost' particle will be the one with the smallest numeral index (up to periodic boundary conditions). For example, if the $m$th particle skipped its neighbor from the right, then $Y_m(t+\dd t)=\min(Y_m(t)+\dd X(t),Y_{m+1}(t))$ and $Y_{m+1}(t+\dd t)=\max(Y_m(t)+\dd X(t),Y_{m+1}(t))$.


We consider two initial conditions. (A) Uniform distribution, where we first pick $M$ positions $X_{m,\mathrm{a}}(0)\sim\mathrm{U}(0,L)$, which are then stored in increasing order within $\{Y_{m,\mathrm{a}}(0)\}$, \ie $Y_{m,\mathrm{a}}(0)<Y_{m+1,\mathrm{a}}(0)$. The subscript `$\mathrm{a}$' stands for the fact that it would yield the annealed variance. (B) Crystalline configuration, being just a Dirac comb at positions $Y_{m,\mathrm{q}}(0)=L\cdot(m-1)/M$. The subscript `$\mathrm{q}$' reminds that computing the annealed variance with these initial condition coincides with the quenched average for uncorrelated initial conditions~\cite{BJC22,DMS23}. See below and Appendix~\ref{apa}.

For each initial condition and set of parameters $\brho$ and $E$, we repeat the above simulation $S=10^3$ times, labelled by $s\in \{1,\cdots,\ S\}$. We estimate the average displacement of the $m$th particle (the tracer) under relabelling at each time-step from the average
\begin{equation}
    \langle \Delta Y_{m,\alpha}(t)\rangle =\frac{1}{S}\sum_{s=1}^S\left[Y_{m,\alpha}^{(s)}(t)-Y_{m,\alpha}^{(s)}(0)\right].
\end{equation}
where $Y_{m,\alpha}^{(s)}(t)$ is its position at time $t$ for the $s$th simulation of either the uniform (annealed, $\alpha=\mathrm{a}$) or crystalline (quenched, $\alpha=\mathrm{q}$) initial distribution. Its second moment is estimated similarly from
\begin{equation}
\langle \Delta Y_{m,\alpha}^2(t)\rangle =\frac{1}{S}\sum_{s=1}^S\left[Y_{m,\alpha}^{(s)}(t)-Y_{m,\alpha}^{(s)}(0)\right]^2.
\end{equation}
The estimator for the tracer's annealed average is then given by
\begin{equation}
\langle \Delta Y^2_{m,\alpha}(t)\rangle_{\mathrm{ac}} = \langle \Delta Y_{m,\alpha}^2(t)\rangle-  \langle \Delta Y_{m,\alpha}(t)\rangle^2.
\end{equation}
Thus we averaged over realizations for the tracer and, since we chose a different random seed for every run, also over initial conditions for the uniform (annealed) initial condition. Here lies the convenience in estimating the quenched variance from the annealed averaging with crystalline initial conditions; otherwise, we would have needed to run many noise realizations for the same randomly-chosen initial configuration, in addition to picking the initial configurations. 

Now, due to the periodic boundary conditions, the system is completely invariant under rotation, and thus any of the $M$ particles could have been chosen as the tracer. Therefore, so to increase data reliability, we take an ensemble average, and find
\begin{subequations}
\label{simres}
\begin{eqnarray}
    \langle \Delta Y^2(t)\rangle_{\mathrm{ac}}&=&\frac1M\sum_{m=1}^M\langle \Delta Y^2_{m,\mathrm{a}}(t)\rangle_{\mathrm{ac}},\\
    \langle \Delta Y^2(t)\rangle_{\mathrm{qc}}&=&\frac1M\sum_{m=1}^M\langle \Delta Y^2_{m,\mathrm{q}}(t)\rangle_{\mathrm{ac}}.
\end{eqnarray}
\end{subequations}
Note that performing the sum over particles already with the sum over realizations would have given unnecessary (arguably correlated, as they come from different particles for the same realization) cross terms. We now compare the numerical results with our analytical predictions, Eq.~\eqref{predPOT}.

\begin{figure}
    \centering
    \includegraphics[scale=0.9]{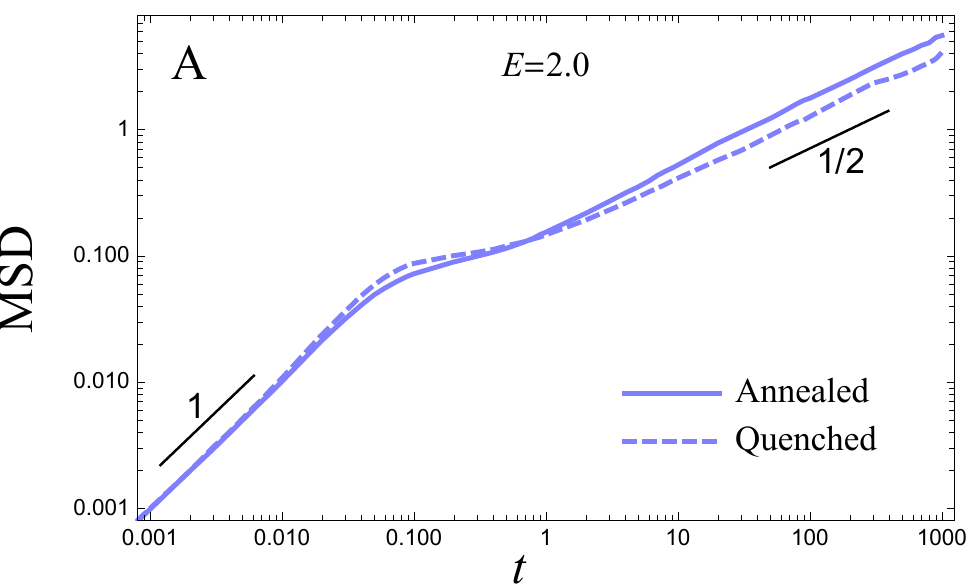}\includegraphics[scale=0.9]{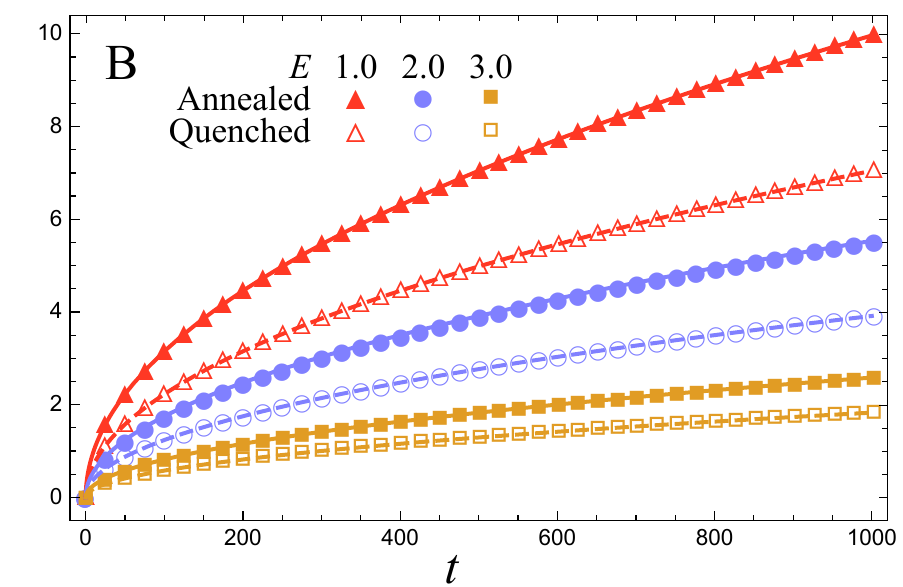}
    \caption{\textit{Numerical Simulation, Periodic Potential}. Tracer's annealed and quenched variances (mean-squared displacement, MSD) versus time, for density $\brho=2$. Panel A: A log-log plot for $E=2.0$, obtained from the simulations. It shows a normal-diffusive regime for short times (no collisions yet and almost-free diffusion around the potential's minima), a brief plateau in between (slow-down due to the potential barriers), and exponent-$1/2$ subdiffusion at late times (after homogenization and sufficient interparticle collisions). Panel B: MSD plots for potential strengths $E$ as indicated. Points are the variances obtained from simulations, while the immediately-adjacent lines are the corresponding theoretical predictions, Eqs.~\eqref{predPOT}.}
    \label{FIGmsdPOT}
\end{figure}

In Fig.~\ref{FIGmsdPOT} we plot the annealed and quenched variances for density $\rho=2.0$. Figure~\ref{FIGmsdPOT}A is a log-log plot, showing in detail the various regimes in typical SFD in a periodic potential, as obtained from the simulations with potential strength $E=2.0$. At first, indeed the particles diffuse normally (with their bare diffusion coefficient, $D_0=1/2$), as they have not encountered their neighbors and typically they have not sampled the potential barriers. Afterwards, we have two simultaneous effects. The more apparent one is the (almost) plateau, which has to do with the particles attempting to overcome the potential barrier, which they succeed doing only with a finite Arrhenius rate (not shown; see Ref.~\cite{SDA22}). Beyond this time, the particles reach their homogenized regime with $\Deff$. The second effect is the collisions with neighboring particles, which causes a crossover from normal diffusion to anomalous subdiffusion with exponent $1/2$. Since $\rho$ and $E$ are of the same order (in fact, equal) in Fig.~\ref{FIGmsdPOT}A, these two effects are not decoupled and occur together in between $t\in[0.1,10]$. Asymptotically, with the above two taking effect, the exponent-$1/2$ subdiffusion occurs with a $\Deff$ equal to the {\em bare} diffusion coefficient. Indeed, this  is what is seen in Fig.~\ref{FIGmsdPOT}B, where all variances grow as $t^{1/2}$. Here, the points shown are the measured variances from the simulations (Eqs.~\eqref{simres}) for different potential strengths $E$ as indicated, while the lines depict the theoretical predictions (Eqs.~\eqref{predPOT}).

\begin{figure}
    \centering
    \includegraphics[scale=0.9]{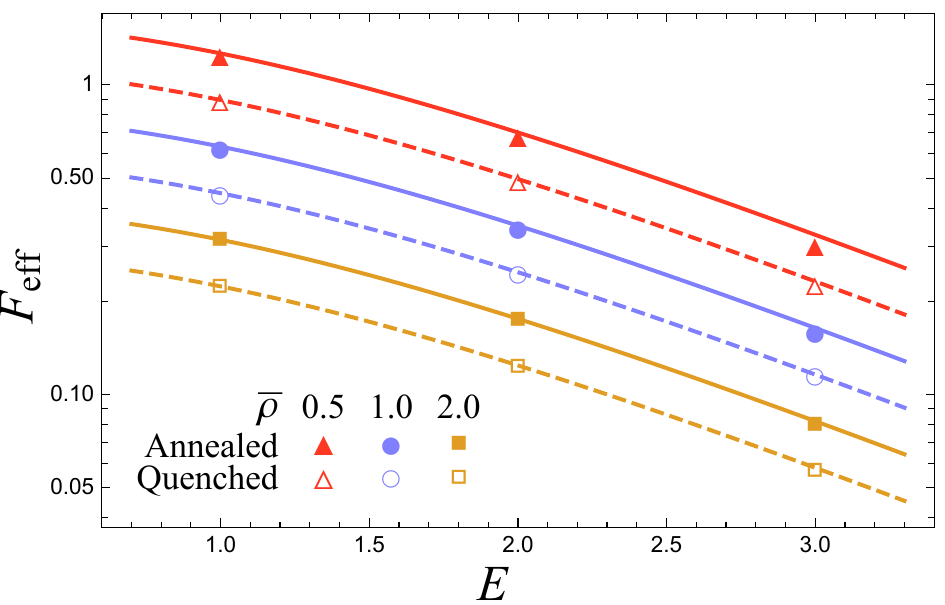}
    \caption{\textit{Numerical Simulation, Periodic Potential}. Tracer's annealed and quenched {\em anomalous} late-time diffusion constant versus potential strengths $E$, for different densities $\brho$, as indicated. Points are the values obtained from simulations using a least-squares linear fit to the variances' log-log plots (as in Fig.~\ref{FIGmsdPOT}A), while the immediately-adjacent lines are the corresponding theoretical predictions.}
    \label{FIGpot}
\end{figure}

In Fig.~\ref{FIGpot} we plot the {\em anomalous} late-time diffusion coefficient, $F_\mathrm{eff}=\langle\Delta Y^2(t)\rangle/t^{1/2}$, which we extracted from a least-squares linear fit in the log-log plots of the variances for points at times $t\in[500,1000]$. We see a good agreement of the simulation with theory overall. The numerical values of $F_\mathrm{eff}$ exhibit relative deviations $\mathcal{O}(1\%)$ from the analytical ones, commonly being an underestimation of the latter. Indeed the errors are most significant in the low density region (since less collisions occur, and thus the approach to the subdiffusive behaviour is delayed) and high energy barriers (as it takes longer to overcome the potential barriers, and hence the {\em homogenization} regime is reached later). This is supported by the fact that the errors get bigger for \emph{all} points when we include MSD values from earlier times (namely, $t\in[100,1000]$; not shown).

\begin{figure}
    \centering
    \includegraphics[scale=0.9]{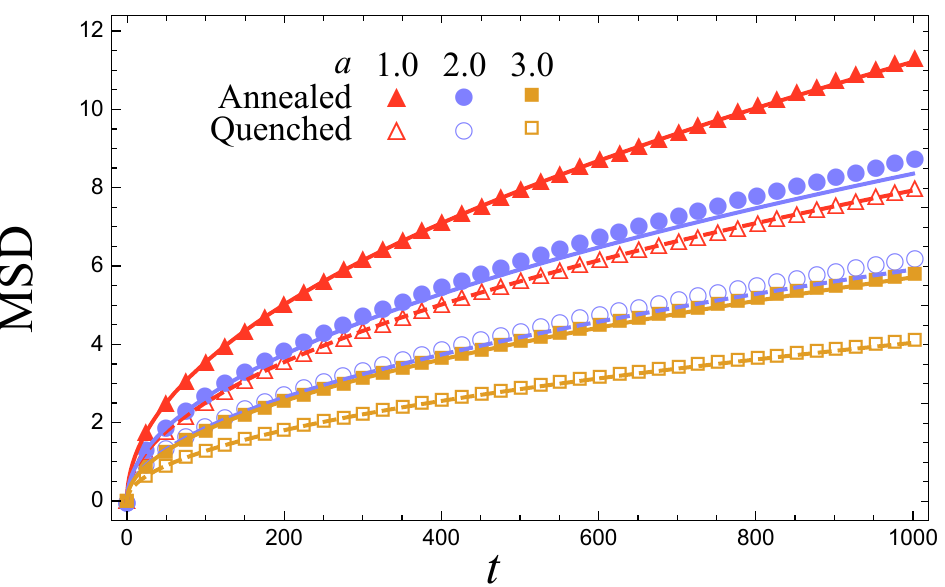}
    \caption{\textit{Numerical Simulation, Periodic Diffusivity}. Tracer's annealed and quenched variances (mean-squared displacement, MSD) versus time, for density $\brho=2$, and diffusivity variabilities $a$ as indicated. Points are the variances obtained from simulations, while the immediately-adjacent lines are the corresponding theoretical predictions, Eqs.~\eqref{predDIF}. For $a=3.0$, a shorter timestep duration was used.}
    \label{FIGmsdDIF}
\end{figure}

\begin{figure}
    \centering
    \includegraphics[scale=0.9]{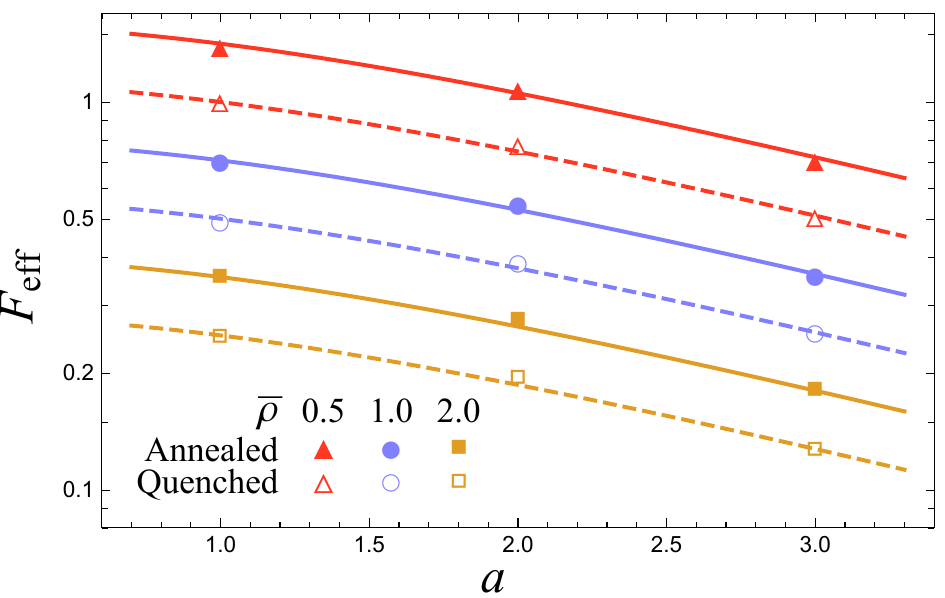}
    \caption{\textit{Numerical Simulation, Periodic Diffusivity}. Tracer's annealed and quenched {\em anomalous} late-time diffusion constant versus diffusivity variability $a$, for different densities $\brho$, as indicated. Points are the values obtained from simulations using a least-squares linear fit to the variances' log-log plots, while the immediately-adjacent lines are the corresponding theoretical predictions. For $a=3.0$, a shorter timestep duration was used.}
    \label{FIGdif}
\end{figure}

To test our prediction further, we also consider the case of periodic diffusivity, given by $D(x)=D_0\exp[a\cos(2\pi x/\ell)]$, in the absence of potential. The effective Lifson-Jackson diffusivity of Eq.~\eqref{lj} is given by $\Deff(a)=D_0/I_0(a)$, and so our predictions for the MSDs read
\begin{subequations}
\label{predDIF}
\begin{eqnarray}
    \langle Y^2(t;a,\rho)\rangle_\mathrm{ac}&=& \frac{2}{\brho}\sqrt{\frac{D_0 t}{I_0(a)\pi}},\\
    \langle Y^2(t;a,\rho)\rangle_\mathrm{qc}&=& \frac{\sqrt2}{\brho}\sqrt{\frac{D_0 t}{I_0(a)\pi}}.
\end{eqnarray}
\end{subequations}
The simulation procedure and the calculation of the MSD are identical to before. (We rely on fewer repetitions of the simulations, $S=200$.) However, due to the exponentiation of the diffusivity variability $a$, the diffusivity is very high at $x=k\ell$ ($k\in\mathbb{Z}$) for big $a$s, and so, for $a=3.0$ we had to reduce the timestep duration to $\Delta t=10^{-4}\ell^2/(2D_0)$. Otherwise, the MSDs are overestimated (as seen, to some extent, for $a=2.0$). With this correction, Fig.~\ref{FIGmsdDIF} shows that the MSDs extracted from the simulation agree well with the theory; so to improve them further, one could consider lower $\Delta t$s, as well as longer-running simulations. A least-squares linear fit in the log-log plot of the MSDs yields the late-time anomalous diffusion constants for this case, which are shown along with the theoretical prediction in Fig.~\ref{FIGdif}.

\section{Conclusions}\label{conc}

We have revisited the question of whether one can use the effective diffusion constant for a single particle in a medium of periodically-varying potential and diffusivity~\cite{TM06} to describe the late-time dispersion of single-file diffusing particles in such media. In agreement with previous simulations~\cite{TM06,LRM19} and the ones of Sec.~\ref{num}, we prove that one can indeed replace the bare diffusivity with the effective diffusion constant. This implies that one may use a coarse-grained description to understand SFD in inhomogeneous systems. Although it would have been very surprising were it not so (by sheer intuition), the demonstration is non-trivial. 

In our scheme, we have adapted the ideas behind the mathematical theory of homogenisation~\cite{PS08,AL92,NG89}, couching them in terms of a late-time asymptotic expansion (more familiar to physicists), which may be of use in other systems. In the context of SFD, it would be useful to rigorously justify a similar utilization of the system-appropriate $\Deff$ in cases such as particles with pairwise additive hydrodynamic correlations and direct interactions~\cite{KOL03} or in the presence of random diffusivity or external potential~\cite{DDH07}. The method we propose rigorously justifies the coarse-graining procedure evoked to derive the macroscopic fluctuation theory~\cite{BDGJL01,BDGJL05,BDGJ06}, whereby short-ranged interactions are incorporated into similar effective diffusion constants and mobilities in a stochastic-hydrodynamic description of interacting particle systems.

\begin{acknowledgments}
We thank Haim Diamant for extensive and enlightening discussions. B.S. acknowledges support from the Israel Science Foundation (Grant No.\ 986/18). D.S.D. would like to thank the Mortimer and Raymond Sackler Institute of Advance Studies in Tel Aviv for their support of his visit to Tel Aviv University where this work was initiated and also acknowledges financial support from the European Union through the European Research Council under EMet- Brown (ERC-CoG-101039103) grant.
\end{acknowledgments}

\appendix

\section{Mimicking quenched variance from a periodic initial distribution}\label{apa}

In this appendix, we give an argument why choosing periodic crystalline initial conditions and computing the resulting annealed variance yields the quenched one. This point is discussed in Refs.~\cite{BJC22,DMS23}. Denote the random initial particle positions that are to the left of the origin as $\{X_m(0)\}$, and define the initial density field as $\rho_\mathrm{L}(x_0)=\sum_{m=1}^M\delta(x_0-X_m(0))$, which is random as well. 
First, we rewrite Eq.~\eqref{npmmnt} as
\begin{equation}
    \langle N^2_+(t)\rangle= \sum_{m=1}^M   \langle \Theta(X_m(t))\rangle + \sum_{m=1}^M\sum_{n=1}^M   \langle \Theta(X_m(t))\rangle \langle \Theta(X_n(t))\rangle -\sum_{m=1}^M  \langle \Theta(X_m(t))\rangle \langle \Theta(X_m(t))\rangle. \label{NpNpbdel}
\end{equation}
Then, we express Eqs.~\eqref{Npbdel} and~\eqref{NpNpbdel} in terms of $\rho_\mathrm{L}(x_0)$ as
\begin{subequations}
\begin{eqnarray}
    \langle N_+(t)\rangle&=&\int_0^ \infty dx \int_{-L}^0 dx_0 p(x,t|x_0,0)\rho_\mathrm{L}(x_0).\\
    \langle N^2_+(t)\rangle&=&\int_0^ \infty dx \int_{-L}^0 dx_0 p(x,t|x_0,0)\rho_\mathrm{L}(x_0)\nonumber\\
    &&+\int_0^\infty dx \int_0^\infty dx' \int_{-L}^0 dx_0 \int_{-L}^0 dx'_0p(x,t|x_0,0) p(x',t|x'_0,0)\rho_\mathrm{L}(x_0)\rho_\mathrm{L}(x'_0)\nonumber\\
    &&-\int_0^\infty dx \int_0^\infty dx' \int_{-L}^0 dx_0 p(x,t|x_0,0) p(x',t|x_0,0)\rho_\mathrm{L}(x_0),
\end{eqnarray}
\end{subequations}
where we kept the (half-) box size $L$ finite for the moment.
Thus, upon taking the average over initial conditions, the general expression for the annealed average in terms of $\rho_\mathrm{L}(x_0)$ is 
\begin{multline}
    \langle N^2_+(t)\rangle_\mathrm{ac}= \int_0^ \infty dx \int_{-L}^0 dx_0\ p(x,t|x_0,0) \overline{\rho_\mathrm{L}(x_0)}+\int_0^\infty dx \int_0^\infty dx' \int_{-L}^0 dx_0 \int_{-L}^0 dx'_0p(x,t|x_0,0) p(x',t|x'_0,0)\times\\
    \times\left[\overline{\rho_\mathrm{L}(x_0)\;\rho_\mathrm{L}(x'_0)}- \overline{\rho_\mathrm{L}(x_0) }\delta(x_0-x_0')-\overline{\rho_\mathrm{L}(x_0)}\; \overline{\rho_\mathrm{L}(x'_0)}\right].\label{dens}
\end{multline}
In the case where the particles are initially independent, only the first term in Eq.~\eqref{dens} contributes, and we recover the annealed result for independent particles\,---\,Eq.~\eqref{n2ac}. 

We have shown in the arguments leading to Eq. (\ref{page}) that quantities of the form
\begin{subequations}
\begin{equation}
\int_{-\infty}^0 dx_0 p(x,t|x_0,0) g(x_0), 
\end{equation} 
can be evaluated at late times as 
\begin{equation}
\int_{-\infty}^0 dx_0 p(x,t|x_0,0)\langle g\rangle_\mathrm{1p},
\end{equation}
\end{subequations}
if $g$ is a periodic function and $\langle g\rangle_\mathrm{1p}$ indicates the spatial average of $g$ over one period. 

In this appendix, we consider a system where the initial positions of the particles are placed on a periodic lattice with randomness given only by global small (a multiple of $\ell_0$) translations. The density is thus random but periodic (\ie hyperuniform). For Eq.~\eqref{dens} it implies (in the limit $L\to\infty$) that
\begin{subequations}
\begin{equation}
\int_0^ \infty dx \int_{-L}^0 dx_0\ p(x,t|x_0,0) \overline{\rho_\mathrm{L}(x_0)} \to \int_0^ \infty dx \int_{-\infty}^0 dx_0\ p(x,t|x_0,0) \overline{\langle\rho_\mathrm{L}\rangle_\mathrm{1p}}
\end{equation}
and
\begin{multline}
\int_0^\infty dx \int_0^\infty dx' \int_{-L}^0 dx_0 \int_{-L}^0 dx'_0p(x,t|x_0,0) p(x',t|x'_0,0)\overline{\rho_\mathrm{L}(x_0)\;\rho_\mathrm{L}(x'_0)}\to \\\int_0^\infty dx \int_0^\infty dx' \int_{-\infty}^0 dx_0 \int_{-\infty}^0 dx'_0p(x,t|x_0,0) p(x',t|x'_0,0)\overline{\langle\rho_\mathrm{L}\rangle_\mathrm{1p}^2}.
\end{multline}
\end{subequations}
This can be seen by carrying out the integrals over $x_0$ and $x_0'$ prior to the average over initial conditions.
The disorder in the above, which can be due to a random lattice translation, is thus irrelevant at late times (as should clearly be the case on physical grounds). This means that the first and third terms in Eq.~\eqref{dens} cancel, yielding
\begin{equation}
\langle N^2_+(t)\rangle_\mathrm{ac}= \int_0^ \infty dx \int_{-\infty}^0 dx_0\ p(x,t|x_0,0) \overline{\rho_\mathrm{L}(x_0)}-\int_0^\infty dx \int_0^\infty dx' \int_{-\infty}^0 dx_0 p(x,t|x_0,0) p(x',t|x_0,0)\overline{\rho_\mathrm{L}(x_0)},
\end{equation}
which corresponds to the quenched variance for uncorrelated particles, as given by Eq.~\eqref{n2qc}.

\end{document}